# Concept Trees: Building Dynamic Concepts from Semi-Structured Data using Nature-Inspired Methods

Kieran Greer, Distributed Computing Systems, Belfast, UK.
http://distributedcomputingsystems.co.uk
Version 1.3

***Abstract*** **–** This paper[1] describes a method for creating structure from heterogeneous sources, as part of an information database, or more specifically, a 'concept base'. Structures called 'concept trees' can grow from the semi-structured sources when consistent sequences of concepts are presented. They might be considered to be dynamic databases, possibly a variation on the distributed Agent-Based or Cellular Automata models, or even related to Markov models. Semantic comparison of text is required, but the trees can be built more, from automatic knowledge and statistical feedback. This reduced model might also be attractive for security or privacy reasons, as not all of the potential data gets saved. The construction process maintains the key requirement of generality, allowing it to be used as part of a generic framework. The nature of the method also means that some level of optimisation or normalisation of the information will occur. This gives comparisons with databases or knowledge-bases, but a database system would firstly model its environment or datasets and then populate the database with instance values. The concept base deals with a more uncertain environment and therefore cannot fully model it beforehand. The model itself therefore evolves over time. Similar to databases, it also needs a good indexing system, where the construction process provides memory and indexing structures. These allow for more complex concepts to be automatically created, stored and retrieved, possibly as part of a more cognitive model. There are also some arguments, or more abstract ideas, for merging physical-world laws into these automatic processes.

---

[1] Accepted by 'Complex system modelling and control through intelligent soft computations' - Book Chapter, by: Studies in Fuzziness and Soft Computing, Springer-Verlag, Germany.

# 1 Introduction

The term 'concept base' has been used previously (Jarke et al., 1995; Zhao et al., 2007, for example) and has been adopted in Greer (2011) to describe a database of heterogeneous sources, representing information that has been received from the environment and stored in the database for processing. The key point is that the information received from one source does not have to be wholly consistent with information received from another source. The uncertain environment in which it operates, means that information can be much more fragmented, heterogeneous, or simply unrelated to other sources. This could be particularly true in a sensorised environment, when sensors provide a relatively small and specific piece of information. As the sensor-based information would be determined by the random and/or dynamic environment in which it operates, there can be much less cohesion between all of the different input sources. For example, event 1 triggers sensor A with value X and shortly afterwards, event 2 triggers sensor B with value Y. Later, event 1 again triggers sensor A with value X, but instead, event 3 occurs and triggers sensor B with value Z. While the nature of the data is much more random, statistical processes can still be used to try to link related pieces of information. This linked data can then represent something about the real world. The term 'concept' can be used to describe a single value or a complex entity equally and so the concept base can consistently store information from any kind of data source. Intelligent linking mechanisms can be used to try to turn the smaller, more simplistic and separate concepts into larger, more complex and meaningful ones. This is probably also more realistic in terms of what humans have to deal with, in our interaction with the real world.

While information might be input and stored in an ad-hoc manner, it is probably the case that some level of structure must firstly be added to the information, before it can be processed, data mined, or reasoned over. When looking for patterns or meaningful relations; then if the data always appears to be random, it is more difficult to find the consistent relations and so a first stage that does this would always be required. This paper looks at a very generic and simplistic way of adding structure to the data, focusing particularly on using whatever existing structure there is, as a guide. Other statistical processes can then use the structure to try to generate some knowledge. Thinking of the sensors or data streams, for example - if it can be determined that concepts A and B usually occur together, while concepts C and D also occur together; knowledge might be able to tell us that when A-B occurs, C-D is likely to occur soon afterwards, or maybe should occur as something else. The current context is to extract this structure from textual information sources, but this is only an example of how the method would work. If consistent patterns can be found, they can be used to grow 'concept trees'. A concept tree is essentially and AND/OR graph of related concepts that grows naturally from the ordering that already exists in the data sources. This paper is concerned with describing the structure of these concept trees and how the process might work. Note that this is at the structure-creation level and not the knowledge-creation level just mentioned.

The rest of this paper is organised as follows: section 2 describes what type of information might be received and why it can be useful. Section 3 gives examples of related work. Section 4 gives step-by-step examples of how the process might work. Section 5 tries to define the processes formally, as would be done for a database. Section 6 gives some suggestions, relating the structure more closely to nature or the physical world. Section 7 describes how this fits in with an earlier cognitive model and linking mechanisms research, while section 8 gives some conclusions on the work.

## 2  Adding Structure to Semi-Structured Data

Computers require some level of standardisation or structure, to allow them to process information correctly. The problem is therefore how to add this structure,to give the computer system a standardised global view over the data. Even the idea of structure is not certain and can be different for different scenarios. Therefore, obtaining the correct structure probably also means the addition of knowledge to the system. As described in the related work in section 3, this type of modelling started with relational databases (Codd, 1970), but then extended to semi-structured and even completely unstructured information. Wikipedia[2] explains that distinct definitions of these are not clear for the following reasons:

---
[2] http://en.wikipedia.org/wiki/Unstructured_data.

1. Structure, while not formally defined, can still be implied.
2. Data with some form of structure may still be characterised as unstructured if its structure is not helpful for the processing task at hand.
3. Unstructured information might have some structure (semi-structured) or even be highly structured but in ways that are unanticipated or unannounced.

The introduction of random events and time elements means that the data sources can also change (Zhang and Ji, 2009), requiring statistical or semi-intelligent processes to recognise patterns that cannot be determined beforehand. This could result in a different type of modelling problem than for a traditional database. For one scenario, the designer creates a model of what he/she wishes to find out about and then dynamically adds specific data instances, as they occur, to try to confirm the model. For another scenario, the actual model itself is not known but is derived from an underlying theory. With the second situation, not only are the model values updated dynamically, but the model itself can change in a dynamic and unknown way.

## 2.1 Types of Data Input

With regard to the text sequences considered in this paper, Greer (2011) describes how a time element can be used to define sequences of events that might contain groups of concepts. A time stamp can record when the concept is presented to the concept base, with groups presented at the same time being considered to be related to each other. This is therefore built largely on the 'use' of the system, where these concept sequences could be recognised and learnt by something resembling a neural network, for example. The uncertainty of the real world would mean that concept sequences are unlikely to always be the same, and so key to the success is the ability to generalise over the data and also to accommodate a certain level of randomness or noise. The intention is that the neural network will be able to do this relatively well. It is also true that there is a lot of existing structure already available in information sources, but it might not be clear what the best form of that is. Online datasets, for example, can be continuous streams of information, defined by time stamps. While the data will contain structure, there is no clearly defined start or end, but more of a continuous and cyclic list of information, from which clear patterns need to be recognised.

As well as events, text might be presented in the form of static documents or papers that need to be classified. For the proposed system, there are some simple answers to the problem of how to recognise the existing structure. The author has also been working on a text-based processing application. One feature of the text processor is the ability to generate sorted list of words from whole text documents. Word lists can also appear as cyclic lists and patterns can again be recognised. This current section of text, for example, is a list of words with nested patterns. In that case, structure could be recognised as a sequence, ending when the word that started the sequence is encountered again. To sort the text, each term in the sequence could be assigned a count of the number of times it has occurred, as part of the sequence. How many times does 'tree' follow 'concept' for example, but a sequence can be more than one word deep. Sequences that contain the same words, or overlap, can be combined, to create the concept trees in the concept base. To select starting or base words, for example, a bag-of-words with frequency counts can determine the most popular ones. The decision might then be to read or process text sequences only if they start with these key words. Pre-formatting or filtering of the text can also be performed. Because this information would be created from existing text documents, the process would be more semantic and knowledge-based. This does not exclude the addition of a time element however and a global system would benefit from using all of these attributes.

The concept trees can then evolve, adding and updating branches as new information is received. Processing just a few test documents however shows that different word sorts of the original data will produce different sequences, from which these basic structures are built, so the decision of correct structure is still quite arbitrary. On the technical front, it might be more correct to always use complete lists of concepts, as they are presented or received, and then try to re-structure the trees that they get added to. Each tree should try to represent some sort of distinct entity and it would be desirable to have larger trees, as this indicates a greater level of coherence. The structure must also be as stable as possible however and so we could try to always add to a base set of concepts, so that the base always has the largest count values. Therefore a triangular structure is realised, with respect to count values, where the base has the largest count, narrowing to the branches. If this basic principle is broken, it might be an indication that the structure is incorrect. Additions to an existing tree should include additions from the base upwards when possible, with new concepts creating new branches if required. It should 'extend' the existing tree along the whole of one of its branches.

## 2.2    Structure Justification

An earlier paper (Greer, 2011) gave a slightly philosophical argument that if two concepts are always used together, then at some level they are a single entity. This is a very general rule not related to any particular application, but describes how any sort of entity can be important based on its relevance to the scenario. Consider then the following made-up scenario: There is a farm with a fence in a field. A sheep comes up to the fence and jumps over it. Sensors in the field record this and send the information to the concept base. The concept base recognises the sheep and the fence objects and assigns them to be key concepts in the event. With our existing knowledge, we would always assign more importance to the sheep, but if we had never encountered either object, maybe the sheep and the fence would be equally important to ourselves as well. The scenario continues, where a cow comes up to the fence and jumps over it, then a chicken comes up to the fence and jumps over it. In this case, the fence now becomes the main and key concept. Without the fence, 'none' of the scenarios can occur. A count of these concepts would give the fence the largest total, again suggesting that it is the key concept. The process to combine these scenarios might then compare these stored events and decide that a concept tree with the fence at its base would be the most stable. This process is described further after the related work section, where the addition of existing knowledge is also suggested, to add a natural ordering to things.

## 3    Related Work

The related work section is able to include topics from both the information processing and AI areas. After introducing some standard data processing techniques and structures, some intelligent methods, relating to nature in particular, are described. It would also be an im-portant topic for problems like data management in the business or online worlds, for example Blumberg and Atre (2003) or Karin (2012). While concepts are the main focus of interest, combining service functionality is more important for the Internet or Cloud at the moment (Aslam et al., 2007; Atkinson et al., 2007, for example). The paper Carr et al. (2001) describes slightly earlier ideas about linked data and marking-up documents on the Internet. It notes how the lines between search and link, or web and database have become blurred and even just searching over metadata tags can be considered as a sort of database operation.

## 3.1 Ontologies and Semantics

A tree structure, or directed graph, is often used to model text sequences, because it allows for the reuse of sequence paths, extending from the same base. Ontologies are essentially definitions of domains that describe the concepts in that domain and how they relate to each other. A section from the book Greer (2008, chapter 4) describes that ontologies can be used to represent a domain of knowledge, allowing a system to reason about the contents of that domain. The concepts are related through semantics, for example, 'a car is a vehicle'. For traditional constructions, relations can then be organised into hierarchical tree-like structures. The 'subclass' relation is particularly useful, where the previous example shows that a car is a subclass of a vehicle. There are different definitions of what an ontology is depending on what subject area you are dealing with. Gruber (1993) gives the following definition for the area of 'AI and knowledge representation', which is suitable for this work:

> 'An ontology is an explicit specification of a conceptualisation. The term is borrowed from philosophy, where an ontology is a systematic account of Existence. For knowledge-based systems, what 'exists' is exactly that which can be represented. When the knowledge of a domain is represented in a declarative formalism, the set of objects that can be represented is called the universe of discourse. This set of objects, and the describable relationships among them, are reflected in the representational vocabulary with which a knowledge-based program represents knowledge. Thus, we can describe the ontology of a program by defining a set of representational terms. In such an ontology, definitions associate the names of entities in the universe of discourse (e.g., classes, relations, functions, or other objects) with human-readable text describing what the names are meant to denote, and formal axioms that constrain the interpretation and well-formed use of these terms.'

This is a desirable definition, but because a concept base is constructed slightly differently, the related ontology construction will also be slightly different. The additional knowledge that defines something like 'subclass' is not automatically present, where the system has to determine the correct position, relation and ordering for any concept, mostly from statistics. Because the knowledge is missing however, the relation must also be more simplistic and would probably normally just be 'related to'. It is also worth noting that the future vision of the Web would probably require distributed ontologies. Again from Greer (2008), the future Internet should maybe describe itself at a local level, with larger centralised representations being created by specific applications, based on the domains of information that they typically use. This would naturally happen as part of the Semantic Web. The construction of these ontologies will enable computers to autonomously search the Internet and interact with the services that are provided and is also part of knowledge management on the Internet. The book 'Towards the Semantic Web: Ontology-Driven Knowledge Management' (2003) discusses the ontology construction problem in relation to p2p networks and the Semantic Web. They note that for reasons of scalability, ontology construction must be automated, based on information extraction and natural language processing technologies. However, for reasons of quality, the process still requires a human in the loop, to build and manipulate ontologies. With a slightly reduced knowledge-level, it is intended that the concept base can construct itself almost completely autonomously, giving it a major advantage in this respect.

For dynamic or autonomic systems, the context in which the knowledge is used can become a critical factor. Context is an information space that can be modelled as a directed graph, rather like an ontology. Context allows both recognition and mapping of knowledge, by providing a structured and unified view of the world in which it operates (Coutaz et al., 2005). It is about evolving, structured and shared spaces that can change from one process to the next, or even through the duration of a single process. As such, the meaning of the knowledge will evolve over time. The key lies in providing an ontological foundation, an architectural foundation, and an approach to adaptation that all scale alongside the richness of the environment. Contexts are defined by a specific set of situations, roles, relations and entities. A shift in context corresponds to a change in the set of entities, a change in the set of possible relations between entities, or a change in the set of roles that entities may play. Unfortunately, the context adaptation cannot currently be carried out in a totally automatic way and a concept base would not really consider context in the first instance. It is constructed primarily through statistical counts, but groups of terms presented at the same time can provide some level of context.

By describing the domain in a standardised way, the programs that use the domain will be able to understand what the domain represents. Through this process, different programs on the Internet will be able to learn about each other and form useful associations with other programs that provide the information that they require. This will enrich the knowledge that they can provide, thus turning the Internet into a knowledge-based system, rather than primarily as a source for direct information retrieval. This is of course, a utopian idea that has many possibilities and may never be fully realised.

## 3.2  Dynamic Databases

As a concept base is a type of database, this is probably the first technology to look at, where the following text is also taken from the book Greer (2008, chapter 3). Databases are the first kind of organised information system, where the first models were developed in the 1960s. The relational model proposed by E. F. Codd (1970) has become the de-facto standard and contains a sound mathematical foundation, allowing for optimisation of the storage and retrieval processes. During the 1980s, research on databases focused on distributed models, in the 1990s object-oriented models and then in the 2000s on XML-based models. The distributed models were necessary because of the evolution of the Internet and networking, which meant that distributed sites of related information could now be linked up electronically. The object-oriented models then arose with the invention of object-oriented programming and the theory that object-based models are a preferable way to store and manipulate information. Then with the emergence of XML as the new format for storing and representing information, XML-based models also needed to be considered. While XML is now the de-facto standard for describing text on the Internet, meaning that most textual information will soon be stored in that format, it has not replaced the relational model for specific modelling. Neither has the object-oriented model. The increased complexity of these models can make them more difficult to use in some cases, when the mathematical foundations of the relational model remains appealing.

### 3.3 New Indexing Systems

The recent problems that 'Big Data' provides, linking up mobile or Internet of Things with the Web, has meant that new database structures, or particularly, their indexing systems, have had to be invented. Slightly more akin to Object-oriented databases are new database versions such as NoSql and NewSql (Grolinger et al., 2013), or navigational databases[3]. As stated in Grolinger et al. (2013), the modern Web, with the introduction of mobile and sensor devices has led to the proliferation of huge amounts of data that can be stored and processed. While the relational model is very good for structured information on a smaller scale, it cannot cope with larger amounts of heterogeneous data. It is usually required to process full tables to answer a query. As stated in Grolinger et al. (2013), CAP (Gilbert and Lynch, 2002) stands for 'consistence, availability and partition tolerance' and has been developed along-side Cloud Computing and Big Data. 'More specifically, the challenges of RDBMS in handling Big Data and the use of distributed systems techniques in the context of the CAP theorem led to the development of new classes of data stores called NoSQL and NewSQL.' They note that the consistency in CAP refers to having a single up-to-date instance of the data, whereas in RDBMs it means that the whole database is consistent. NoSql now has different meanings and might also be termed 'Not Only SQL'. It can use different indexing systems that might not even have an underlying schema. So it can be used to store different types of data structure, probably more as objects than tables. The database aspect however can try to provide an efficient indexing system, to allow for consistent search and retrieval over the distributed contents. There are different data models for implementing NoSql. 'Key-value stores' have a simple data model based on key-value pairs, which resembles an associative map or a dictionary. The key uniquely identifies the data value and is used to store and retrieve it from the data store. The data value can be of any type. In 'column-family stores' the data are stored in a column-oriented way. One example might be where the dataset consists of several rows, each of which is addressed by a unique row key, also known as a primary key. Each row is composed of a set of column families, and different rows can have different column families. Similarly to key-value stores, the row key resembles the key, and the set of column families resembles the value represented by the row key. However, each column family further acts as a key for the one or more columns that it holds, where each column consists of a name-value pair. 'Document stores' provide another derivative of the key-value store data model by using keys to locate documents inside the data store. Each document can be highly heterogeneous and so the store can provide the capability to index also on the document contents. 'Graph databases' originated from graph theory and use graphs as their data model. By using a completely different data model to the other 3 types, graph databases can efficiently store the 'relationships' between different data nodes. Graph databases are specialized in handling highly interconnected data and therefore are very efficient in traversing relationships between different entities. NewSql is based more on the relational model, where clients would interact in terms of table and relations. Its internal data model however might be different and there can be semi-relational models as well.

---

[3] http://en.wikipedia.org/wiki/Navigational_database.

A navigational database is a type of database in which its records or objects are found primarily by following references from other objects. Navigational interfaces are usually procedural, though some modern systems like XPath (XPath, 2014), can be considered to be simultaneously navigational and declarative. Navigational databases therefore use a tree indexing system and can fall under the graph-based category of NoSql. These graph-based databases therefore look more similar to a concept base or concept tree. While the problems of semi-structured or unstructured data remain, these new databases do offer general architectures and indexing systems. One criticism of graph-based ones however, is that they tend to lead to very messy sets of indexing links that do not have very much structure. This is possible for concept trees as well, but as the concept tree might have a more specific construction process, it can provide some kind of mathematical foundation to help with the organisation.

## 3.4 Semantic Environment

As well as a sensorised environment, a concept base is also closely related to the Web 3.0, that is, the Semantic Web (Berners-Lee, Hendler and Lassila, 2001) combined with Service Oriented Architectures (SOA) (OASIS, 2014). This is because they can also produce individual pieces of semantic information dynamically and computer-to-computer processing likes to link these up. This would mean that real-time information retrieved from sensors, for example, could be combined with more knowledge-intensive, but static information provided by the Internet, to answer a wider variety of queries. A hierarchical structure is also appealing for reasons of organisation and search efficiency, and so as has been suggested previously by other researchers (Robinson and Indulska, 2003), at least a shallow hierarchy would be useful. The largest network of information that we have at the moment is of course the Internet. This is composed of many individual Web sites that contain information by themselves. However, the only relation to other Web sites is through hyperlinks that are typically created by human users. This is really the only way to try and combine the information provided into a meaningful whole. To try and turn the Internet into a network of knowledge, the Semantic Web has thus been invented. With the Semantic Web, the programs that run on the Internet can describe themselves through metadata, which will allow other programs to look them up and be able to understand what they represent. Metadata is 'data about data' and provides extra descriptive information about the contents of a document or piece of information. If this information is available in a machine-readable format, then computer-to-computer interaction will be enabled as well as the typical human-to-computer interaction.

While the Internet is the main source for information, an evolving area is that of mobile devices, including the Pervasive sensorised (Hansmann, 2003) or Ubiquitous computing (Greenfield, 2006) environments. The mobile environment, by its very nature, is much more dynamic. The Internet contains static Web pages that once loaded will remain on a server, at a site from where they can be located. With mobile networks, devices may be continually moving and so they may connect and disconnect to a network at different locations. Ubiquitous computing is a model of human-to-computer interaction in which information processing has been integrated into everyday objects and activities. An example of this would be to embed sensors into our clothes, to identify us when we went to a particular location. This dynamism actually presents problems to a network that tries to organise through experience. The experience-based organisation requires some level of consistency to allow it to reliably build up the links, but if the structure constantly changes then this consistency may be lost. However, the mobile devices may be peripheral to the main knowledge content. They would be the clients that want to use the knowledge rather than the knowledge providers. For example, in the case of people wearing sensors, it would be the building that they entered that would learn from the sensor information and provide the knowledge, not the people themselves. The sensors would continually be bringing new information into the environment that would need to be processed and integrated. The paper Encheva (2011) also includes the ideas of concept stability and nesting, which are central to the whole problem. The following sections describe how the laws of nature have helped with building these complex structures.

### 3.5 Underlying Theories and the Natural World

If the model cannot be pre-defined, then it needs to be learned. To do this, the computer program needs to be given a set of rules to use as part of the construction process. For a generic solution, these rule sets are usually quite simplistic in nature. Again taken from Greer (2008, chapter 1), Complex Adaptive Systems is a general term that would also comprise the sciences of bio-inspired computing. The term Complex Adaptive Systems (or complexity science), is often used to describe the loosely organised academic field that has grown up around the study of such systems. Complexity science encompasses more than one theoretical framework and is highly interdisciplinary, seeking the answers to some fundamental questions about living, adaptable and changeable systems. A Complex Adaptive System (for example, Holland, 1995; Kauffman, 1993) is a collection of self-similar agents interacting with each other. They are complex in that they are diverse and made up of multiple interconnected elements and adaptive in that they have the capacity to change and learn from experience. One definition of CAS by John Holland (1995), one of the founders of this science, can also be found in Waldrop (1993) and is as follows:

'A Complex Adaptive System (CAS) is a dynamic network of many agents (which may represent cells, species, individuals, firms, nations) acting in parallel, constantly acting and reacting to what the other agents are doing. The control of a CAS tends to be highly dispersed and decentralised. If there is to be any coherent behaviour in the system, it has to arise from competition and cooperation among the agents themselves. The overall behaviour of the system is the result of a huge number of decisions made every moment by many individual agents.'

The nature of the interactions between the individual entities is the key aspect that distinguishes such complex systems from complicated systems (Al-Obasiat and Braun, 2007). A system is called complex if the interactions between its components are not predictable and if it has at least one or more of the following characteristics:

- It is non-linear.
- It is dynamic.
- It is time-variant.
- It is chaotic or stochastic.

All telecommunication networks possess one or more of these attributes. Complicated systems are an alternative type of complex system. However, while complicated systems interact in a predictable way, with CAS, the unpredictable interactions between individual components in the system give rise to 'emergent' behaviour. Emergence is the process of complex pattern formation from simpler rules. An emergent behaviour arises at the global or system level and cannot be predicted or deduced from observing the behaviour of the individual components in the lower-level entities. Because of external forces, concept trees would probably be classified as complex, because their construction is unpredictable.

### 1.1.1 Mathematical Theories

If one considers the natural world, then Cellular Automata might be thought to be relevant and at some level they provide the required mechanisms. There are different versions of Cellular Automata (Wolfram, 1983, for example). They work using a localised theory and entropy (Shannon, 1948) could be a key consideration for the structure that is described in the following sections. As described in Wikipedia[4]: In thermodynamics (Rudolf Clausius, 1862), entropy is commonly associated with the amount of order, disorder, and/or chaos in a thermodynamic system. For a modern interpretation of entropy in statistical mechanics, entropy is the amount of additional information needed to specify the exact physical state of a system, given its thermodynamic specification. If thought of as the number of microstates that the system can take; as a system evolves through exchanges with its environment, or outside reservoir, through energy, volume or molecules, for example; the entropy will increase to a maximum and equilibrium value. The information that specifies the system will evolve to the maximum amount. As the microstates are realised, the system achieves its minimum potential for change, or best entropy state. In information theory, entropy is a measure of the uncertainty in information content, or the amount of unpredictability in a random variable (Shannon, 1948). As more certainty about the information source is achieved, the entropy (potential uncertainty) reduces, to a minimum and more balanced amount.

However, it would be difficult to map these types of state machine, or mini-computers, over to a process that is designed only to link up text, to create ontologies. Most distributed systems use some kind of localised theory as well, in any case. The reason for this section is the fact that the dynamic linking uses a basic association equation to create links and also, as described later, makes a decision about breaking a link and creating a new structure. To show their relation to distributed systems and nature, the following quote is from the start of the paper (Wolfram, 1983).

> 'It appears that the basic laws of physics relevant to everyday phenomena are now known. Yet there are many everyday natural systems whose complex structure and behaviour have so far defied even qualitative analysis. For example, the laws that govern the freezing of water and the conduction of heat have long been known, but analysing their consequences for the intricate patterns of snowflake growth has not yet been possible. While many complex systems may be broken down into identical com-ponents, each obeying simple laws, the huge number of components that make up the whole system act together to yield very complex behaviour.'

---

[4] http://en.wikipedia.org/wiki/Entropy, plus_(information_theory), _(statistical_thermo-dynamics), or _(order_and_disorder), for example.

If we know what the underlying theory of the system is, then it can build itself in a distributed manner, even if we do not know what the eventual structure will be. Cellular Automata would be too rigid for a concept tree, as they can be created from a fixed grid structure with local interactions only; while a concept tree is required to create and move structures, as well as link up existing ones. Fractals (Mandelbrot, 1983; Fractal Foundation, 2014) are also important and cover natural systems and chaos theory. There are many examples of fractals in nature. Using a relatively simple 'feedback with minor change mechanism', the complex systems that actually exist can be created. As described in Wolfram (1983), automata and fractals share the feature of self-similarity, where portions of the pattern, if magnified, are indistinguishable from the whole. Tree and snowflake shapes can be created using fractals, for example. Fractals also show how well defined these natural non-bio processes are already. So automata would belong to the group called fractals and are created using the same types of recursive feedback mechanism. The construction of a concept tree would be a self-repeating process, but the created structures are not self-similar. However, they would result from same sort of simplistic feedback mechanism that these self-similar systems use.

Agent-Based modelling is another form of distributed and potentially intelligent modelling. Scholarpedia[5] notes that Agent-Based Models (ABM) can be seen as the natural extension of the Ising model (Ising, 1925) or Cellular Automata-like models. It goes on to state that one important characteristic of ABMs, which distinguishes them from Cellular Automata, is the potential asynchrony of the interactions among agents and between agents and their environments. Also ABMs are not necessarily grid-based nor do agents 'tile' the environment. An introduction to ABM could be the paper Macal and North (2006). Agent-based models usually require the individual components to exhibit autonomous or self-controlled behaviour and to be able to make decision for themselves, sometimes pro-actively. While Cellular Automata would be considered too inflexible, agents would probably be considered as too sophisticated. Although as noted in Macal and North (2006), some modellers consider that any individual component can be an agent (Bonabeau, 2001) and that its behaviour can be as simple as a reactive decision.

---

[5] Scholarpedia http://www.scholarpedia.org/article/Agent_based_modeling.

### 1.1.2 Biologically-Related

As Artificial Intelligence tries to do, there are clear comparisons with the natural world. Comparisons with or copying of the biological world happens often, but trying to copy the non-biological world is less common, at least in computer science. There are lots of processes or forces that occur in the non-biological world that have an impact on physical systems that get modelled. Trying to integrate, or find a more harmonious relationship between the two, could be quite an interesting topic and computer programs might even make the non-bio processes a bit more intelligent. It might currently have more impact in the field of Engineering and the paper Goel (2013) describes very clearly how important the biological designs are there. With relation to a concept base, a small example of this sort of thing is described in section 6. As noted in Wolfram (1983) and other places, as the second law of thermodynamics implies, natural systems tend towards maximum entropy, or a minimal and balanced energy state. While biological ones tend towards order as well, the non-biological ones tend towards disorder. Cellular Automata are therefore closer to biological systems, where a cross-over is required if non-biological systems are going to exhibit the same levels of intelligence. Although, something like wave motion in the sea shows a steady and consistent behaviour until the wave breaks. So the equations of wave motion can certainly work together in a consistent manner. Even the snowflake shows consistent behaviour for its growth stage, and so on. So with the non-biological systems, a consistent energy input can be the controlling mechanism that triggers specific mechanics. If this is lost or changes, the system can behave more chaotically and may have problems healing or fixing itself. Biological systems might be driven by something more than the specific mechanic, which allows for another level of - possibly overriding - control. Consider the re-join (self-heal?) capability of the concept tree later, in sections 4 and 5.

The Gaia theory (Lovelock and Epton, 1975) should probably be mentioned. As stated in Wikipedia: 'The Gaia hypothesis, also known as Gaia theory or Gaia principle, proposes that organisms interact with their inorganic surroundings on Earth to form a self-regulating, complex system that contributes to maintaining the conditions for life on the planet'. So the inorganic elements of the planet have a direct effect on the evolution of the biological life. Maybe the inorganic mechanisms suggested here are for smaller individual events, than the more gradual self-regulation of a large global system. Although in that case, they could still be the cause for global changes.

## 4     Concept Tree Examples

The examples provided in this section show how the concept trees can be built from text sequences. They also describe some problems with the process and the proposed solutions. With these examples, it is more important to understand the general idea than consider them as covering every eventuality. Section 5 then tries to give a more formal definition, based on these examples. To show how a tree is created, consider the following piece of text: *The black cat sat on the mat. The black cat drank some milk.* If punctuation and common words are removed, this can result in the following two text sequences:

*Black cat sat mat*
*Black cat drank milk*

From this, as illustrated in Figure 1, a tree can be built with the following counts. The base set of 'black cat' can be extended by either the set 'sat mat' or the set 'drank milk'. The base 'black cat' concept set has been found from a sort that starts with these terms, where combining the two sets of terms then reinforces the base. It also appears when constructing these trees that sets of counts should in fact balance, unless additions with missing information are allowed. The counts for the immediate child nodes should add up to the count for the parent. If, for example, a new list starting with 'sat mat' was allowed to be added, it would only increment counts higher up the tree, altering the tree's balance. If this caused the triangular rule to be broken, a re-structuring process, starting where the count becomes larger again should 'prune' the tree and create a new one, with the more stable branch at its base. As will be suggested in the rules of section 5, in fact, if the trees are always constructed from the base up, this particular problem will not exist.

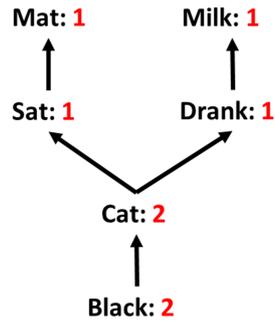

Figure 1. Concept tree generated from two text sentences.

### 4.1 Combining on Common Branches

If the following text was also stored in the concept base: *The thirsty boy drank some milk. The thirsty elephant drank some milk.* This could result in two more concept sequences:

*Thirsty boy drank milk*

*Thirsty elephant drank milk*

To add these to the concept base data structure, the process might firstly create two new trees, one starting with 'thirsty boy' and another with 'thirsty elephant'. However, the terms 'drank milk' have now become the most important overall and therefore should be at the base of a tree. Adding to Figure 1, the 'drank milk' branch of the 'black cat' tree should be pruned and added with the other two 'drank milk' sequences, to start a new tree, with a count of 3, as shown in Figure 2. It would then be necessary to add links between the trees, where they were related. Links can be created using indexing or unique key values, for example. The structure of each tree can then develop independently and so long as they exist, any links between them will remain, giving some level of orderly navigation. So why separate the concepts and not just extend the 'black cat' tree? One reason is the new base concept sets of 'thirsty boy' and 'thirsty elephant', creating a new tree by default. Separating and linking is also an optimisation or normalisation feature. If the process concludes that 'drank milk' is an important concept in its own right, it should not be duplicated in different places, but rather it should be stored and updated in one place and referenced to by other trees. The understanding of 'drinking milk' is the same for all 3 animals. Then of course, also the triangular count rule.

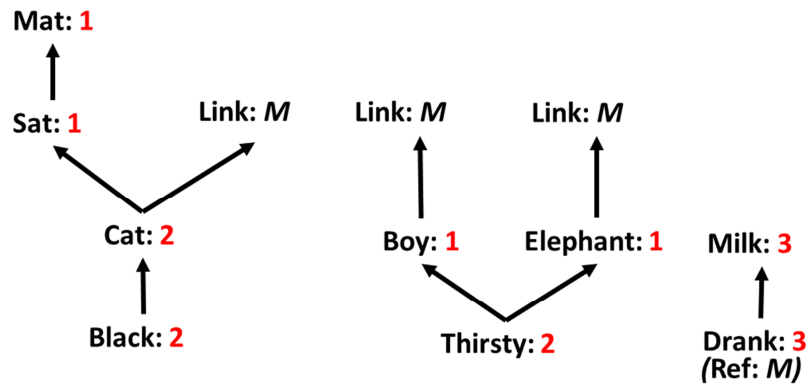

Figure 2. Example of 'pruning' and optimising the structure when new trees are added.

## 4.2 Combining with Unrelated Branches

Another situation would be if the concept base then receives, for example, 3 instances of: *The thirsty elephant drank milk and ate grass*. This would automatically add 'ate grass' to the 'drank milk' tree and the count would still be OK. The sequence 'ate grass' however, only relates to the original elephant branch in this case. There is no indication that the boy or cat ate grass. The final solution to this is another new tree and also a new indexing key, to link the elephant with both the 'milk' and the 'grass' trees. This might happen however after a stage of monitoring (see section 4.3). The other tree branches keep the original index, where Figure 3 shows what the new set of structures might look like. Note the way that existing links only need to be traversed if the parent tree also has the key value; and it is more important to note the types of structure and links that get created, rather than the exact semantics of the whole process. For example, start with the first linking keyset, remove a link key if it gets used and only continue if the next link key is in the list.

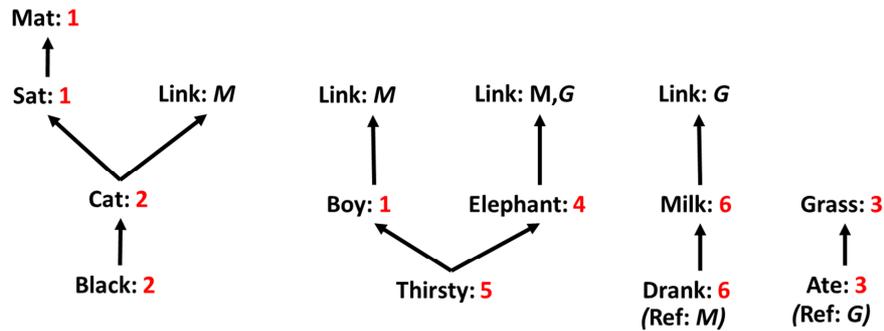

Figure 3. New tree and key value changes the indexes.

Not shown in the figure, the key sets might possibly be similar to the NoSql column-family stores, with a primary key retrieving a set of secondary keys. The secondary keyset would then allow for the traversal of the linked database trees. The primary key can relate to an external entity that is interested in certain groups of concepts. As each tree is quite separate, the secondary keys could, for example, relate or point to sets of base tree concepts. In this case, graph-like navigation is also possible, as leaf nodes can link to other tree base nodes as well, where the secondary keyset helps to define the allowed starting paths. So this is another normalising possibility, but would require that entities followed similar rules in general. If the primary key defines a pool of secondary keys that define what trees can be traversed, then using certain primary keys for particular events will again force some normalisation, as almost similar keysets might get combined.

### 1.1.3 Query Example

A query process is still required to access the tree contents and can be related to the keysets that exist. For example, if the cat also starts to eat grass, a primary key that points to the 'Thirsty' tree, might then include the 'Black' tree in its secondary keyset as well. The 'Cat' link now also includes the G graph link. Without any additional information, the traversal can return either the cat or the elephant for milk and grass. This might need to be specified, because another similarly indexed primary key could accept, cat, boy or elephant for just the milk concept. It might also be interesting to consider that the secondary keyset allows for the completion of circuits through the graph trees. For example, if the first primary key also includes the 'Ate' tree (and maybe 'Drank') in its secondary keyset and requires all indexed trees to be true; then only cat and elephant can be returned. This is just a possibility and is described further in section 7.2. It does however look like useful searches might require a set of conditions, but also allow for some automatic reasoning, where a query language is possible. For this example, it looks like a Horn clause could be used (Jarke et al., 1995; Greer, 2011, for example).

## 4.3 Compound Counts

Another requirement is to break the structure up again. If the concept base, for example, received concept sequences of 'drank milk with a long trunk', with no relation to any of the animal tree base concepts; then the process should just add 'long trunk' to the 'drank milk' base. This is not necessarily incorrect because that specific information does not state that the cat or the boy do not have long trunks. If the information is then used and the cat or boy trees traversed, it will incorrectly return that they have long trunks. Some type of compound count can be used to check for this. A *positive:negative* compound count, for example, can indicate that the tree is possibly incorrect. If the negative count becomes too large, then the 'drank milk long trunk' tree can be split and a link added between them instead. The new link key could then get added to some primary keysets, to allow for its traversal. The trick is in being able to determine when the information is untrue. Does the information need to be returned and then evaluated as untrue, or can the update process allow for this to be recognised automatically? For example, a new text sequence of 'thirsty boy drank milk' is again added to the database, where it updates the related tree nodes. As it stops short in the 'drank milk' branch, any nodes further up that branch can have their negative count incremented instead. As the elephant tree reinforces the positive count here however, this is then an indication that the tree should be split, as the information is true sometimes and false other times. A break in any tree would automatically create a new indexing key as well. This would be sent to all related trees that can then decide what link keyset best relates to them – the one for just 'drank milk' or the one that also links to 'long trunk'. New entries can therefore be flagged in the first instance, until they become reliable.

There is also a process of reasoning and adjustment here, again over a period of time. Even if a tree branch is not proven to be incorrect, in some scenarios, a negative count might be required to indicate that part of a tree is no longer relevant to the current entity or scenario. More traditionally, a decay factor can be used to determine this. If, for example, the link is rarely used, its value decays until it is lost completely. If it is used so infrequently, then it might as well not be present, even if it is not false. So this is another alternative link update mechanism that could be added, but a compound key helps to decide to split rather than remove completely. With a single value that gets incremented or decremented, you have to judge how many times each has occurred. If there is a compound count, then this is clear and it is easy to tell if the branch is true as well as false.

### 4.4 Re-Join or Multiple References

A final consideration might be the re-joining of one tree to another and also a problem with multiple reference links to the re-joined part. As the data can be random, it might initially be skewed in some way and force a tree to break into two trees. Over time this evens out and the original single tree can become correct again. This is determined by the fact that the counts become consistent with a single tree structure again. In that case, it would be possible to re-join the previous branch that is now a base, back onto the first tree. The only worry would be if there are also multiple references to the second tree that had a branch broken off. These references might not relate to the original tree as well. It might not be good practice to allow arbitrary references half-way up a tree and so if the previous branch has a different set of references now, then maybe it must stay as a base. Ideas here would therefore include transferring back only some of the new tree, while keeping the rest as the second tree, with its base. The next section tries to explain this again, but more formally. This might be the most 'intelligent' part of the process, as a re-join can be compared to a self-heal or fixing process. The non-bio systems would typically not do this and therefore continue to a more chaotic state.

## 5      Formal Specification

The concept tree idea, for a concept base, has a restricted construction process. It is based on a frequency count with a very strict rule about relative count sizes. It might therefore be possible to define the construction process more formally and even bring some standardisation or normalisation; where other similar techniques, such as Navigational Database or NoSql, are not able to. The following sets of declarations might be useful to standardising the process and bring added order to the structures. Initial tests have confirmed some of these rules, but are not variable enough to test all of the possible scenarios. Because the rules are more of a logical argument than a mathematical proof, they are listed as groups of points.

### 5.1    General

1. A concept tree can represent different types of entity. The entity however should be a whole concept. For example, it might be a single object in the real world, or a single action. Therefore, the base concepts in any tree are the ones that would be used first in any scenario.
2. Tree structures require that every child node has a count that is the same or less than its parent. This should always be the case if the linking integrity is maintained, unless branches are allowed to re-join.
3. Whenever possible, the process would prefer larger trees for the following reasons:
    a. A larger tree has more meaning as a general concept and gives added confidence when reasoning over its group of nodes.
    b. A larger tree gives more coherence to the concept base.
    c. Larger trees mean less of a trend towards a chaotic structure.
4. Normalisation would like each concept to exist only once and so, also for this reason, the whole process tries to find what the main concepts are and place them as base concepts to trees. As with traditional databases, if a concept exists somewhere only once, then it only needs to be updated in one place. This is difficult or even impossible however, for every scenario:
    a. If the concept gets used for different contexts, then its meaning and relation to other concepts changes slightly, when it needs different link sets.

b. For a distributed system over a large area, it might simply not be practical to have the concept at one place only and be able to find it.

   c. Even if trees have similar branches, a link might be required if other factors do not allow a join.

5. Indexing and linking can use key sets, but it can also include graph-based navigations. This is because the structure is tree-based, with links between tree nodes defining concept sequences.

**5.2   Truth Tests**

1. For a tree to exist, every node in it must be true. That does not mean that every node is evenly used, but there should be no false information. This extends to being true for any entities that link to the tree or related sub-tree. If any part is false for any linking entity, then the tree needs to be split.

2. Note the difference between a part of a tree that is rarely used and a part or path that is false or untrue. Rarely used is OK, but untrue is not.

3. A set of links to a tree from different entities might make parts of the tree untrue, when it then needs to be split at the false branch. It might however be quite difficult to determine if a path is untrue, as information retrieval scenarios might mean that the path simply does not get traversed. So the type of count key can be important and the trick is to be able to recognise when tree paths are rarely used, or are simply false.

   a. There could be a time-based degradation of a link, for example. If it degrades so much as to remove it, then it has never been used and so it is not relevant, even if it is not false.

   b. Or possibly the counting mechanism's 'group:individual' count (Greer, 2011), that reinforces a count, both for the concept group and the individual. This can determine when individual nodes no longer appear to be the same as others in the group.

   c. Or there is a 'positive:negative' count, when the negative count can become too large.

   d. There could also be a time-based count that measures when events happen at the same time. This is important for recognising when trees can be re-joined.

## 5.3 Tree Comparisons

1. Tree comparisons and updates are made using groups of concepts that represent individual input events. The event group is considered to be a complete entity itself and gets stored in the concept base as that. It is then also compared with the other structures as that, where it needs to match with existing tree paths in one of two ways:
    a. If it matches exactly from the base of another tree up any branch, then it can be added to that tree.
    b. If its' base matches to a different node of another tree, then a link between the two trees can be created.
2. If a smaller independent entity is added as a branch to a larger one, then it will not be possible to access it without going through the larger entity first. This means that the normal process of reconstruction will be to break at a tree branch and move to a tree base, with the other direction being used less often.
3. Re-structuring will therefore also prefer to link between trees than to re-join them permanently. This is because a link provides the appropriate navigation, while the base nodes still remain for each tree, allowing them to be accessed directly.
4. While linking is more practical, coherence would prefer permanent joins to create larger trees and so under the correct conditions, a join should be preferred, where any doubt would lead to a dynamic link instead. The re-joining process requires more intelligence, which may be why it would be a more difficult automatic process.

## 5.4 Linking or Joining

1. Any reinforcement of an existing tree, based on adding a new group of concepts, should always start from the base node.
    a. If it would start part of the way up a tree, then the process should form a new tree instead.
    b. Similarly, when a new path is added to an existing tree, it must start from the base only and traverse through any sub-path up to any leaf node. It can then extend that leaf node if required.

2. For a linking example, if we have two trees – tree 1 and tree 2, then:
    a. The tree 2 is simply added as is, with a possible node link to tree 1 and subsequent events can try to change or combine the structures further.
    b. This would be more in line with the general theory, but the idea of entropy or concept coherence would prefer the next scenario if possible.
3. For a permanent join example, if we have two trees – tree 1 and tree 2, then:
    a. The tree 1 can be split at the node related to tree 2's base node and that branch combined with tree 2 for the new structure.
    b. Less likely is a permanent join the other way, but it is still possible. For example, if tree 2 has a path from the base up that matches to a branch in tree1. Then if the counts are OK, the path can be moved from tree 2 to tree 1.
4. Linking related nodes is always possible. Different keysets can then define, for different entities, whether they traverse all of the links or not.
5. Re-joining trees needs to consider the base entity links more.
    a. Breaking off a branch from tree 1 to join to tree 2 at its base would be easier.
        i. If tree 2 has all of the entity links that tree 1 has, then the join can be automatic.
        ii. If tree 2 has additional entity links, then a compound count can be added, because it might still be unclear if the branches, new to some entities, are false.
        iii. If the broken branch however is completely contained in the tree 2 path, then the join can be automatic.
    b. Re-joining tree 2, or part of tree 2 from its base, to a tree 1 branch is more difficult.
        i. If tree 2 does not have any entity links to its base, then it can be added to any other matching branch.
        ii. If tree 2 has a different set of entity links, then its base must be accessible and so it cannot be removed.
        iii. If tree 1 and tree 2 have the same set of entity links, then a join should be attempted. A check might be performed to determine if the two trees are always accessed together. If that is the case, then they can be joined over some common branch or node.

6. Unclear is when one branch has additional elements inside of it, so that the branch it is being compared with, would need to be extended internally and not at an edge. This is quite common with ontologies, for example. This might favour breaking the larger branch at the two points where the new nodes exist, creating 3 trees and linking between them. The first tree uses only two of the new trees while the other uses all 3.
7. A truth test might check if a join is preferable to a link, including branches not defined as false, but possibly now out of character and can be moved.
8. So there could be a statistical, or even a reasoning process that decides what join action to take and this could be different for different implementations.

# 6 Relation to Nature

This section gives some more comparisons with natural laws and is about trying to justify the proposed construction mechanism, by showing that it will give the best possible balance to a concept tree, with the minimum amount of additional intelligence or knowledge required. It is reasonable to think that in the random or chaotic world that we live in, there is no reason to always link from a larger 'measurement' to an equal or smaller one. This is however the main rule of the concept tree and there is some mathematical justification or foundation to it. Some of the evidence was found after the creation of the concept tree, more than the concept tree has been derived from it. However, if it can be used to support the general model or theory, then why not specify it here. The main point to note is the fact that base concepts should probably be the most frequently occurring ones statistically. That is probably a sound enough idea, based on statistics alone. If trying to compare to a real-world physical law, then if tree branches were allowed to become larger again, the tree would probably break at that place. This might be an opportunistic statement, but it is completely the idea behind the triangular counting rule. Other pieces of evidence that might provide support are listed in the following sections.

## 6.1 Problem Decomposition

Any sub-entity must be smaller than the entity it belongs to. This is particularly relevant to the process of problem decomposition that is used to solve large and difficult problems. The larger problem is broken down into smaller ones, until each smaller problem is simple enough to be solved. So this is another application of the natural ordering. It is also the case that you cannot be a sub-concept of something that does not exist. If thinking about Markov models, then one construction of these will count the number of occurrences, of transitions from one state to another. This process will necessarily require the 'from' state to exist first and therefore, if the model is tree-like without loops, each parent state must have a larger or the same count value as the following state, as part of the same rule. Similar to concept trees, Markov models have been used for text classification or prediction, as well as state-based models.

### 6.2   Clustering and Energy

Some of the research that has looked at clustering processes, for example, the single link theory (Sibson, 1973) might provide support. This original theory proposed that any node should link to its closest neighbour. These small clusters could then link to their nearest neighbours in the next iteration, and so on. Therefore, through only one link from each group, at each iteration, larger clusters can eventually be formed. It is interesting to note that if there is a certain ordering of the nodes, this process will work particularly well. A measurement of closeness depends on what is being measured and also the evaluation critera. However, suppose that spatial distance is the metric, where a line of evaluated nodes can only cluster with the node on either side – necessarily being the closest nodes. Consider the two sets of nodes, represented by Figure 4 and Figure 5. In these figures, each node value is represented by its height in the graph and each node position in the cluster space, is represented by its position in the graph.

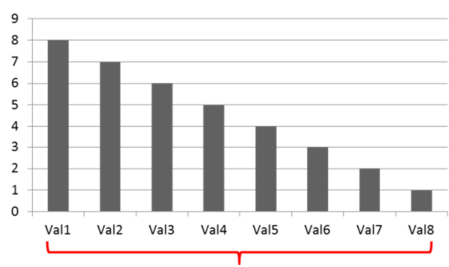

Figure 4. Energy of 7 required to traverse all elements.

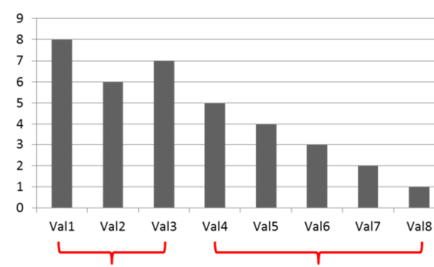

Figure 5.  A single change increases the energy amount to 9.

In Figure 4, the nodes have a perfect descending order. Using the process of linking to your closest neighbour, this could lead to the whole set of nodes creating a single cluster in one go. In Figure 5, the perfect ordering is broken, where node 3 will link with node 2 only. This forces two clusters to be formed, or forces a break in the sequence. We also have the idea of a minimal energy, or entropy (Shannon, 1948). This has already been used to cluster or sort text documents, for example Decision Trees (Quinlan, 1986) and the principle of entropy can also be applied to a concept tree. If one considers the simplistic sorting mechanism in Figure 4 again, it can be seen that the most efficient sort, causing the least amount of energy to move from one place to the next, is in fact the uniform decreasing of the entity lengths, from largest to smallest. If each energy change is 1 unit, then a total of 7 units are required. Any change in this order, for example Figure 5, would require a larger amount of energy to traverse all of the entities – 9 units in this case. As natural systems like lower energy states, a self-organising system might favour the lower energy state. This therefore supports the idea of not adding larger counts to smaller ones, because the required energy amount for the same entity set increases, as in Figure 5. It could increase and then decrease uniformly, but in general, it would support the rule. Entropy also deals with the problem of micro-states, where possibly Figure 4 has only one and Figure 5 has two, but Figure 1 is better because the whole dataset is more coherent and it is already at its minimal state.

### 6.3 Language Structure

Another comparison should be with how we process natural language. Language is so fluent that there are not many restrictions on what can be said or written. As the concept tree is a simplified model of natural language however, it might allow some rules to be included. Generic or autonomous rules are desirable and also plausible. They might be thought of as an extra layer above the basic statistical counts that help to direct the initial structure. They would not be allowed to override the triangular count rule however. The ordering used in WordNet[6] (Fellbaum, 1998; Miller, 1995), for example, is the sort of ordering that would be useful. The base, for example, could typically be formed from nouns and verbs, with adjec-tives or adverbs forming mostly the leaf nodes or end branches. In a real-world sense, the descriptive words would possibly define specific instances of the more grounded noun or verb concept groups. For example, 'the black cat sat on the mat', gives a count of 1 initially to each concept and so the ordering before adding to a tree could be changed. The rule might state to add 'cat' at the base instead of 'black', as it is an object. Then possibly some sort of reverse polish notation 'cat - mat - sat' to push nouns down, or just 'cat - sat - mat'. So the exact language structure might get lost, but the associations will still exist and the rules will help to reconstruct text sequences from the tree. As another example, we can have a cat and a mat, but maybe only the 'black' cat sat on a 'red' mat, and so on. Descriptive nodes at the end would also help to relate the concept tree more closely with earlier work, as described in section 7.

---

[6] I have to note my recent interest in WordNet, although, most of the new theory here was formulated before that, with WordNet then supporting it.

### 6.4 Natural Weight

The following, associated with size or weight, is possibly even more interesting. It would be a strange way of looking at ordering text, but it is again a physical-world rule being applied in a slightly different context and again relates to the idea of sub-concepts. Note that text often relates to real world objects and so its construction would have to be consistent with the physical world. In the real world, it is often the case that the largest and therefore heaviest entity, resides at the bottom of things. Putting a heavier object on-top of a lighter one is not often done and so there is a natural order here. It might be possible to use this knowledge, as part of the tree structure, without requiring more sophisticated natural language or ontology understandings. For example, every event that takes place, takes place on planet earth. If we were creating a structured ontology, planet earth would be at the bottom. Then, for example, a car always drives on a road and so a road should link to a car, not the other way around. It might be the case that the car branch, when it gains relations to lots of other things, would be broken off to form a new base, but it still makes more sense to link from road to car and not car to road. So this ordering, based on some knowledge of relative size or use in the real world, might also become part of a structuring rule. It would be useful because the related context-specific information should not be very sophisticated and so it might be possible to apply the knowledge automatically again. We just need to know that there is a car and a road, for example. One could imagine a large database that stores different bands of entities, grouped simply by size or weight, that are not allowed to be ordered before/after another entity. It is not a typeOf or subClass relation, but a more functional one. Maybe something like relative use, but it is really only the ordering that is required. This fixed ordering would again be a secondary aid, where the statistical counts and dynamic relations of the parsed text would still have the most influence. The trees of Figure 1 to Figure 3 might have their ordering changed slightly, for example, but the word groups and concept associations would still be determined by the dynamic text, not fixed knowledge. For example, the mat should probably be placed before the cat, when the cat branch could be broken off later. It might be 'mat – cat – black + sat', or something.

## 7  Relation to Earlier Work

This section is slightly different, looking at a specific cognitive model, rather than general theories. It is helpful for developing that cognitive model further and will hopefully add ideas for a more intelligent system, but can be skipped if the database model is specifically of interest. Earlier research by the author has looked at how a whole cognitive model might be developed from very simple mechanisms, such as stigmergic or dynamic links (Greer, 2013b; Greer, 2008). The earlier work described how a reinforcement mechanism can be used to determine the reliability of linked source references in a linking structure. These links are created through user feedback only and are therefore very flexible, as the feedback can be much more variable than static rules can accommodate. User feedback adds the intelligence of the user, which a rule set might not contain. While concept trees are also built from user feedback, they are then constrained by pre-determined rules and knowledge. They are also more semantic, complementing the event instances of the earlier work. A concept tree could therefore be created from similar source types – sensor-based, dynamic input, specific concepts, but deal more with the existing structure than the events that created it. It is still possible to make comparisons with earlier work on a neural network model (Greer, 2011) that clustered without considering semantics, but blindly presumed that input presented at the same time must simply be related. The original cognitive model did include an ontology or knowledge-base reference, to provide this type of support. Some comparisons with bio-related models in general can also be made.

## 7.1 Biological Comparisons

More recent work again (Greer, 2013a) has put Hebb's well-known theory of 'neurons that wire together, fire together', into a computer model. It has added a mechanism using the idea that when they fire together they may be attracted to each other and grow links to join up. The rules of section 5.4 (maybe point 5.b) fit in well with this as it suggests comparing the sets of input links to trees. If both trees have the same set of input links, then when these fire, both trees will be activated and can therefore decide to join up. It does not however suggest exactly how they might grow towards each other or combine in a biological model. The earlier cognitive model (Greer, 2013b; Greer, 2008) defines a 3-layer architecture, where the bottom level links for optimisation purposes, the middle layer links to aggregate these pattern groups, while the top layer links to create higher-level concepts and trigger dynamic events. As the concept trees theory is more about aggregating and balance, over all of the data, it is more suited to this middle level. It has also been noted in the formal specification of section 5 that a concept might be duplicated, simply because of the distributed nature of the system. This is also the case for the human brain, as it is known to duplicate information and the practical aspects of trying to access a particular brain region might make it easier to simply duplicate some information locally. Ideas of entropy and automatic monitoring can also be related to both the stigmergic/dynamic linking model (Greer, 2008; section 8.5) or the concept trees. As either system develops, it will tend towards some sort of fixed structure. This trend would then only be broken by a change in the input state. So, after the formations are created, any more dramatic changes might indicate a change in data, and so on. This idea probably applies to most dynamic systems with similar designs.

## 7.2 Higher-Level Concepts

There is a reference to a linking structure in Greer (2008, section 9.3.7, figure 24) that describes how linked concepts might only be related or activated if they are assigned specific values. For example, if we have mother, son and uncle concepts linked, then it might only be true if the mother is called Susan, the son is called John and the uncle is called David. The idea of pushing the descriptive text to the leaf nodes, so as to represent specific instances, has been written about in section 6.3. There is also a reference to another linking structure in Greer (2008, section 9.3.7, figure 25) that tries to index different concept sets through unique keys. It has the same idea as the indexing system being used here and a diagram of this is shown in Figure 6.

The nodes are meant to represent concepts and groups of them, higher-level concepts. However, because there can be overlap between concepts they can be grouped together, with different indexes defining each exact group. If concept trees were used, a tree consisting of A-B-C-D could link to a tree consisting of E only, for example. The ABCD tree would have a base node with some value and then branches, one of 'A to B' and one of 'C to D'. An event entity would then need to activate the base node of the tree and activate all of its branches, to realise the first concept group. To realise the second group, a different event entity would need to link to and activate both the ABCD tree and the E tree, at the same time. Then possibly and interestingly, can a link between the two trees themselves complete a circuit, to indicate the other concept group of CDE. If a link between the leaf D node and the base E tree node exists, for example? This might be a more dynamic model than the original design of Figure 6 that considered fixed sets of (unique) reinforced links only. The key sets possibly sit on-top of the linking structure, where both can change dynamically. So there are two different possibilities for dynamic change, but with the new functionality, there are also other technical difficulties. The intention is that concept groups will represent something more meaningful and therefore can be used as part of a reasoning process. This paper would suggest that it is more of a memory structure, but with the same goal of defining higher-level concepts more accurately, to allow them to be reasoned over.

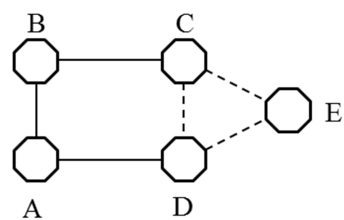

Figure 6. Example network with two higher-level concepts A-B-C-D and C-D-E (Greer, 2008).

## 7.3 Complementary Structures

Dynamic links have therefore been used previously (Greer, 2011; Greer, 2008) as part of a neural network, but the two techniques are probably compatible. It is curious that the knowledge-based concept tree, in relation to the 3-level cognitive model described in Greer (2013b or 2008, section 9.3.8), would be more closely associated with the first optimising level and the second aggregating level. It would create the base or bed of the system. The experience-based neural network would then be more closely associated with the third level. It would manipulate the knowledge (cleverly) through a dynamic, experience-based approach. Looking at the concept trees has actually helped to create a clearer picture and provide some more consistency over the whole model. If the concept trees are used to create pattern groups in the middle level, then it makes sense for them to have a main base concept that defines the tree, with branches or sub-groups that define its contents. It also makes sense for the construction process to start at the general base node and work through to smaller and more specific details at the leaf nodes. It also makes sense that it is more knowledge-based. The earlier neural network model (Greer, 2011) also creates a hierarchical structure, but it was noted that the construction process there might start with the leaf or individual nodes that are then aggregated together into a main or higher-level concept. That neural network model was associated more with the third level of the cognitive model that deals more with dynamic events and triggers. If the concept groups there are based on events, then it could make sense that a reader of those would receive input as small events instances in time. Each event could be some knowledge, defined by some structure. The events would then be aggregated together into something more singular and maybe even learned. They are based on time and external forces, where learning and predicting is also important. But this then gives more sense to the architecture overall and allows for the two hierarchical construction directions to be OK. In a general sense, we already know this. As stated in Greer (2008, section 4.8), with regard to service-based networks: different industries would prefer either a top-down or a bottom-up approach to organisation. Top-down starts with a central component and then adds to it when required. Bottom-up starts with simpler components and then combines them to provide the more complex organisation. If you want more control then a top-down approach is preferred. If you allow a more chaotic but independent organisation, then maybe bottom-up is pre-ferred. It is the same argument for the cognitive model. Top-down relates to knowledge-based concept trees and in this context also, to small but specific entities. Bottom-up relates to the event-based clustering and also to self-organising these smaller structures. As an example, you could imagine a human seeing a tree and learning about its different components or varieties; but when out walking on a stormy day, learning in a different way to avoid falling branches when under a tree in high winds. Or following earlier papers' food examples, you could imagine a human tasting different food types and learning what they are made of; but when in a restaurant, selecting a menu based on the food types and discovering some new recipe through the experience.

## 8  Conclusions and Future Possibilities

This paper has introduced two new ideas of concept trees and concept bases. The concept base is a more general device that is the storage program for the trees. It is also responsible for sorting or creating the trees, and for managing the index and link sets. The concept trees are described in more detail and even formally defined. The counting rule that is introduced in this paper and probably a different construction method, make the concept tree a bit different to other graph-based techniques. The addition of some rules helps to standardise the construction process and give it some mathematical foundation. The idea of only allowing a narrowing structure with respect to count values is probably a good one, because it is statistically consistent and also consistent with the real world. Ideas from nature or the physical world support this and are interesting, but should probably not be taken too seriously. They could introduce a very light form of intelligence, although a light form of knowledge is required first. Any concept is allowed to be a main one and this is defined by an automatic count. The rule set can then give additional structure independently, but it is still the presented data that determines what trees get built. Problems with the process might include the creation of a long list of very short trees that represent nothing in particular by themselves. This then begins to look a bit like the standard memory storage on a computer, with pointers between pieces of memory linking them up. There is however the possibility of building larger more meaningful trees as well. A comparison, or relation, with Markov mod-els has been introduced because they are known to work well and may exhibit the same statistical counting property.

The construction process builds hierarchies automatically and these can represent any type of concept. A slightly weaker idea is therefore to try to build service-based business processes or compositions in the same way, where the earlier stigmergic links were suggested for the same task. See, for example (Greer, 2008; section 7.3.2.1), or maybe (Aslam et al., 2007) or (Atkinson et al., 2007). While real-world concepts or natural language might be restricted by sets of relations that can justify the triangular counting mechanism, more complex business processes might not be. There is a difference between a sub-process and linking two independent processes. In that case, statistical counts would be used purely for reliability, but it is a known problem and several solutions that are at least semi-automatic, have already been suggested. It is worth noting that the count values could be used as probability values, or something similar, as each tree is a bit self-contained. If a particular structure was presented to a network and one of the concepts was missing, the system could try to calculate a probability value, indicating the confidence that the missing concept was in fact an error. This would be an automatic way to assign a value range to the stored data, for security reasons, or other. So concept trees can also be looked at in terms of automatically creating process hierarchies and really does span from the large Internet-based network to the small cognitive model.

Not every group of concepts should be added either and dynamic factors like reinforcement and time can also be considered. So while the construction process is automatic, a reasoning component might also make certain decisions. For example, does a link to another newly created tree actually apply to my instance? If a real tree is taken as the natural world model, and why not, then it obeys the rule that a heavier branch will cause a lighter one to snap. The new AI part then is the idea of an intelligent indexing and linking system, to keep consistency between the split trees. This means that even if the original structures disintegrate, while the natural world entity would tend to chaos, the linked elements will allow for traversal through specific channels and maintain the order. The question would be how efficient or accurate the structure can be.

The idea to use this as part of an indexing and memory structure is optional, but it would fit in well with the cognitive model written about in earlier papers (Greer, 2013a; Greer, 2013b; Greer, 2011; Greer, 2008). The whole process could mostly be performed automatically, with minimum existing knowledge. The earlier model diagrams are relevant enough to be compared with the concept tree directly and even compliment it. This research is still a work in progress and the hope is to be able to provide more substantive results in the future.

## Disclosure

This paper is an updated version of a paper called 'Concept Trees: Indexing and Memory from Semi-Structured Data', originally published on DCS and Scribd, June 2012.